\documentclass[aps,prl,twocolumn,superscriptaddress,notitlepage,nofootinbib,preprintnumbers]{revtex4-1}
\usepackage{amsmath}
\usepackage{amssymb}
\usepackage{color}
\usepackage{xcolor}
\usepackage{graphicx}
\usepackage[colorlinks=true,linkcolor=blue]{hyperref} 
\usepackage{textcomp}
\usepackage{xspace,slashed}
\usepackage[utf8]{inputenc}
\usepackage{subcaption}
\usepackage{multirow}
\usepackage{nccmath} % \mfrac
\usepackage{slashed}
\usepackage{orcidlink}
\usepackage{epsfig}

\begin{document}

\title{
Single-inclusive hadron production in electron-positron annihilation at next-to-next-to-next-to-leading order in QCD
}

\author{Chuan-Qi He~\orcidlink{0009-0004-0737-6518}}
\email{legend_he@m.scnu.edu.cn}
\affiliation{State Key Laboratory of Nuclear Physics and Technology, Institute of Quantum Matter, South China Normal University, Guangzhou 510006, China}
\affiliation{Guangdong Basic Research Center of Excellence for Structure and Fundamental Interactions of Matter, Guangdong Provincial Key Laboratory of Nuclear Science, Guangzhou 510006, China}
\affiliation{Key Laboratory of Atomic and Subatomic Structure and Quantum Control (MOE), Guangdong-Hong Kong Joint Laboratory of Quantum Matter, Guangzhou 510006, China}

\author{Hongxi Xing~\orcidlink{0000-0003-3503-7084}}
\email{hxing@m.scnu.edu.cn}
\affiliation{State Key Laboratory of Nuclear Physics and Technology, Institute of Quantum Matter, South China Normal University, Guangzhou 510006, China}
\affiliation{Guangdong Basic Research Center of Excellence for Structure and Fundamental Interactions of Matter, Guangdong Provincial Key Laboratory of Nuclear Science, Guangzhou 510006, China}
\affiliation{Southern Center for Nuclear-Science Theory (SCNT), Institute of Modern Physics, Chinese Academy of Sciences, Huizhou 516000, China}

\author{Tong-Zhi Yang~\orcidlink{0000-0001-5003-5517}}
\email{tongzhi.yang@m.scnu.edu.cn}
\affiliation{State Key Laboratory of Nuclear Physics and Technology, Institute of Quantum Matter, South China Normal University, Guangzhou 510006, China}
\affiliation{Guangdong Basic Research Center of Excellence for Structure and Fundamental Interactions of Matter, Guangdong Provincial Key Laboratory of Nuclear Science, Guangzhou 510006, China}
\affiliation{Physik-Institut, Universit\"at Z\"urich, Winterthurerstrasse 190, 8057 Z\"urich, Switzerland} 

\author{Hua Xing Zhu~\orcidlink{0000-0002-7129-6748}}
\email{zhuhx@pku.edu.cn}
\affiliation{School of Physics, Peking University, Beijing 100871, China} 
\affiliation{Center for High Energy Physics, Peking University, Beijing 100871, China} 

\preprint{ZU-TH 21/25}

\begin{abstract}
Single-inclusive hadron production in electron-positron annihilation (SIA) represents the cleanest process for investigating the dynamics of parton hadronization, as encapsulated in parton fragmentation functions. In this letter, we present, for the first time, the analytical computation of Quantum Chromodynamics (QCD) corrections to the coefficient functions for SIA at next-to-next-to-next-to-leading order (N$^3$LO) accuracy, achieving the highest precision to date for hadron production processes. Utilizing the BaBar measurement as a benchmark, we assess the phenomenological implications of this high-precision calculation. Our findings demonstrate a substantial reduction in scale uncertainties at N$^3$LO and offer an improved description of the experimental data compared to lower-order calculations. This advancement underscores the importance of higher-order corrections in achieving a more accurate understanding of hadronization processes.
\end{abstract}

\maketitle

%======================================================================
\noindent{\it Introduction.--}
The investigation of hadron production in high-energy particle collisions, such as electron-positron, electron-proton, and proton-proton scatterings, offers profound insights into the non-perturbative dynamics of Quantum Chromodynamics (QCD)~\cite{ParticleDataGroup:2024cfk}. Among these processes, single-inclusive hadron production in electron-positron annihilation (SIA) is particularly significant, as it represents the cleanest and most direct probe of parton hadronization. This hadronization process is quantitatively described by parton fragmentation functions (FFs)~\cite{Berman:1971xz,Field:1977fa,Collins:1989gx}, which encapsulate the probability distributions for quarks and gluons to transition into observable hadrons. FFs are indispensable tools for unraveling the structure of matter and elucidating the intricate dynamics of QCD at low energy scales, making them a fundamental quantity of modern particle physics research~\cite{Albino:2008gy,Metz:2016swz}.

Precision calculations of partonic coefficient functions for hadron production in perturbative QCD (pQCD) are crucial for extracting reliable information about the fundamental QCD properties, such as FFs~\cite{Albino:2008gy} and the strong coupling constant~\cite{Dissertori:2015tfa}, from experimental data. These coefficient functions have been computed up to next-to-next-to-leading order (NNLO) for SIA~\cite{Rijken:1996ns,Mitov:2006wy}, and more recently for electron-proton~\cite{Bonino:2024qbh,Goyal:2023zdi,Bonino:2024wgg,Goyal:2024tmo,Goyal:2024emo} and proton-proton collisions~\cite{Czakon:2025yti}. Progress has also been made toward the NNLO computation of fiducial cross sections with identified hadrons~\cite{Gehrmann:2022pzd,Bonino:2024adk,Czakon:2025yti}. These results, together with the NNLO time-like splitting functions~\cite{Chen:2020uvt,Mitov:2006ic,Moch:2007tx,Almasy:2011eq} has enabled global analyses of fragmentation functions (FFs) at NNLO accuracy~\cite{Bertone:2017tyb,Anderle:2015lqa,Soleymaninia:2020bsq,AbdulKhalek:2022laj,Borsa:2022vvp,Li:2024etc,Gao:2025hlm}. These advances underscore the critical synergy between fixed-order NNLO calculations and global QCD analyses.

NNLO calculations have already led to substantial improvements in reducing theoretical uncertainties and enhancing agreement with experimental measurements. However, given the precision of modern SIA datasets (e.g., BaBar, Belle, LEP, BESIII) and the proposed future lepton colliders, such as the CEPC~\cite{CEPCStudyGroup:2018rmc,CEPCStudyGroup:2018ghi,CEPCStudyGroup:2023quu}, STCF~\cite{Achasov:2023gey}, ILC~\cite{Behnke:2013xla,Bambade:2019fyw,ILCInternationalDevelopmentTeam:2022izu}, and FCC-$ee$~\cite{FCC:2018byv,FCC:2018evy,FCC:2018vvp}, theoretical uncertainties now risk limiting the statistical power of global FF analyses. Achieving N$^3$LO accuracy for SIA is therefore essential to further constrain parton fragmentation functions and advance our understanding of hadronization dynamics. 

In this letter, we present a significant advancement in this direction: we extend the analytic continuation in Ref.~\cite{Chen:2020uvt} to the scattering process in full QCD and, for the first time, obtain the complete results for SIA coefficient functions at N$^3$LO accuracy. To assess the phenomenological impact of our results, we compare our N$^3$LO predictions with the BaBar measurement, a benchmark dataset in the study of SIA. Our findings reveal a substantial reduction in scale uncertainties at N$^3$LO and demonstrate an improved description of the experimental data compared to lower-order calculations. By providing a more accurate and reliable description of parton hadronization, our results contribute to a deeper understanding of the strong interaction.

\noindent{\it Kinematics of SIA.--}
We focus on the photon exchange process of SIA in $e^+ e^-$ collisions
\begin{align}
e^+  \, + e^- \to \gamma^* (q) \to h (p) + X\,,
\end{align}
where $q$ and $p$ represent the four-momenta of the virtual photon and the identified hadron $h$, respectively, $X$ denotes any inclusive final hadronic state. In this process, the unpolarized differential cross section can be written as~\cite{Nason:1993xx,Webber:1994zd} 
\begin{align}
\label{eq:siaDiff}
    \frac{d^2 \sigma^h}{d x \, d\cos \theta} = \frac{3}{8} (1+ \cos^2 \theta) \frac{d \sigma_T^h}{dx}+ \frac{3}{4} \sin^2\theta \frac{d\sigma_L^h}{dx}\,, 
\end{align}
where the variable $\theta$ denotes the angle of $h$ with respect to the electron beam direction in the center of mass frame. The Bjorken variable is defined as
$x = 2 p \cdot q/Q^2$ with $Q^2 = q^2$.
According to the leading twist QCD factorization, the transverse ($T$) and longitudinal ($L$) differential cross sections can be factorized into the following form:
\begin{align}
\label{eq:factorization}
    \frac{d\sigma_k^h}{d x } = \sum_q \sum_j \left(D_{j}^h \otimes C_{j q} + D_{j}^h\otimes C_{j \bar{q}} \right)(x) \, \times\sigma^{(0)}_{q\bar{q}}\,,
\end{align} 
where $k=T,L$, and we omit the index $k$ in the coefficient function $C$. $\sigma^{(0)}_{q\bar{q}}=\frac{4\pi \alpha_e^2}{3Q^2}N_ce_q^2$ is the born-level cross section. The convolution is defined as
$(D \otimes C )(x) = \int_x^1 \frac{dz}{z} D(z) C(x/z)\,.$ In Eq.~\eqref{eq:factorization}, $D_{j}^h$ is parton fragmentation functions, representing the probability density of parton $j$ fragmenting into hadron $h$, and $j$ runs over all parton flavors including gluon. Using charge conjugation invariance of the strong interactions, we have 
$C_{qq} = C_{\bar{q} \bar{q}}, \, C_{\bar{q} q} = C_{q \bar{q}}, C_{g q } = C_{g \bar{q}}\,. $
Then Eq.~\eqref{eq:factorization} can be reorganized into the following form,
\begin{equation}
\label{eq:factorizationSNS}
\begin{aligned}
    \frac{d \sigma_k^h}{d x} &=  D_{g}^h \otimes \left(\textcolor{black}{2 C_{gq}}\right) \sum_{q=1}^{N_f} \sigma_{q \bar{q}}^{(0)}  \\
    &\quad + \sum_{q=1}^{N_f} \left(D_{q}^h + D_{\bar{q}}^h \right) \otimes \left(\textcolor{black}{C_{qq}^\text{V} + C_{q \bar{q}}^{\text{V}}}\right) \sigma_{q\bar{q}}^{(0)} \\
    & \quad + \sum_{q^\prime=1}^{N_f} \left( 
 D_{q^\prime}^h + D_{\bar{q}^\prime}^h\right) \otimes \left( \textcolor{black}{ C_{q^\prime q} + C_{q^\prime \bar{q}} } \right) \sum_{q=1}^{N_f} \sigma_{q\bar{q}}^{(0)}\,,
\end{aligned}
\end{equation}
where $N_f$ is the number of massless quark flavors, and the valence coefficient functions are defined by $C_{qq}^{\text{V}} = C_{qq} - C_{q^\prime q},\,  C_{q \bar{q}}^{\text{V}} = C_{q \bar{q}} - C_{q^\prime \bar{q}}\,.$ In this letter, we determine the N$^3$LO corrections, i.e. the coefficient of $a_s^3$ with $a_s = \alpha_s/(4 \pi)$, to the partonic coefficient functions $C_{gq},\, C^{\text{V}}_{qq} + C^{\text{V}}_{q\bar{q}},\, C_{q^\prime q} + C_{q^\prime \bar{q}} $ in Eq.~\eqref{eq:factorizationSNS}, and we work in dimensional regularization with the dimension $d = 4 - 2\epsilon$.

\begin{figure}[ht!]
    \centering
    \begin{subfigure}{0.48\linewidth}
        \centering
        \includegraphics[width=\linewidth]{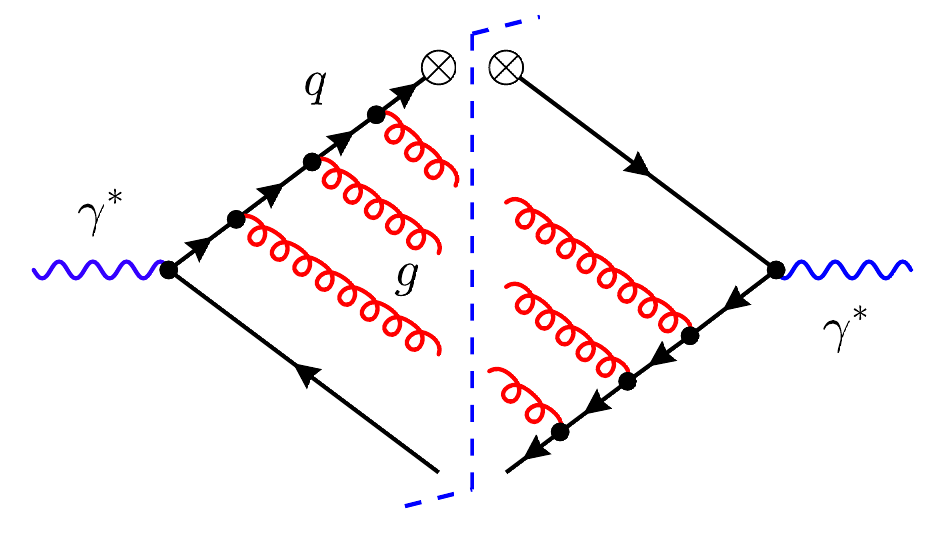}
        \caption{RRR}
        \label{fig:subfig1}
    \end{subfigure}
       \hfill
    \begin{subfigure}{0.48\linewidth}
        \centering
        \includegraphics[width=\linewidth]{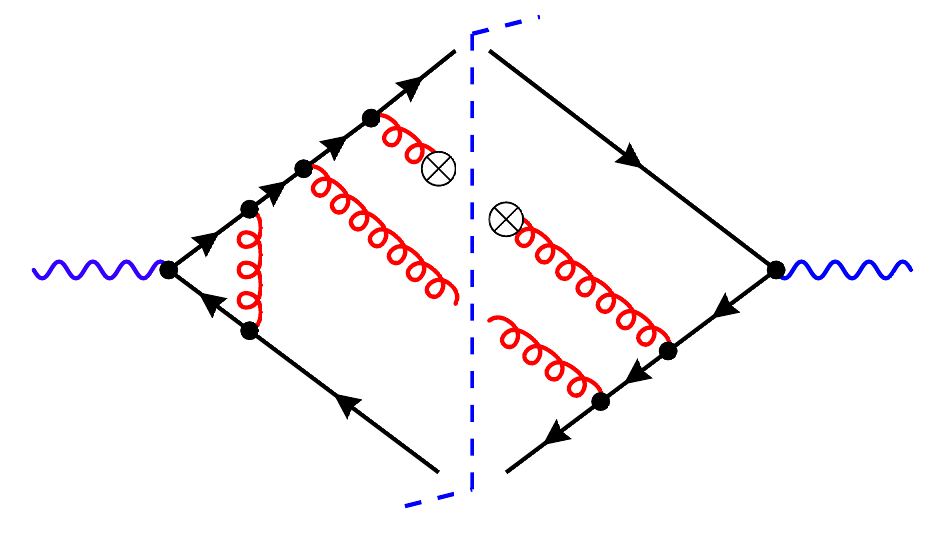}
        \caption{VRR}
        \label{fig:subfig2}
    \end{subfigure}
     \begin{subfigure}{0.48\linewidth}
        \centering
        \includegraphics[width=\linewidth]{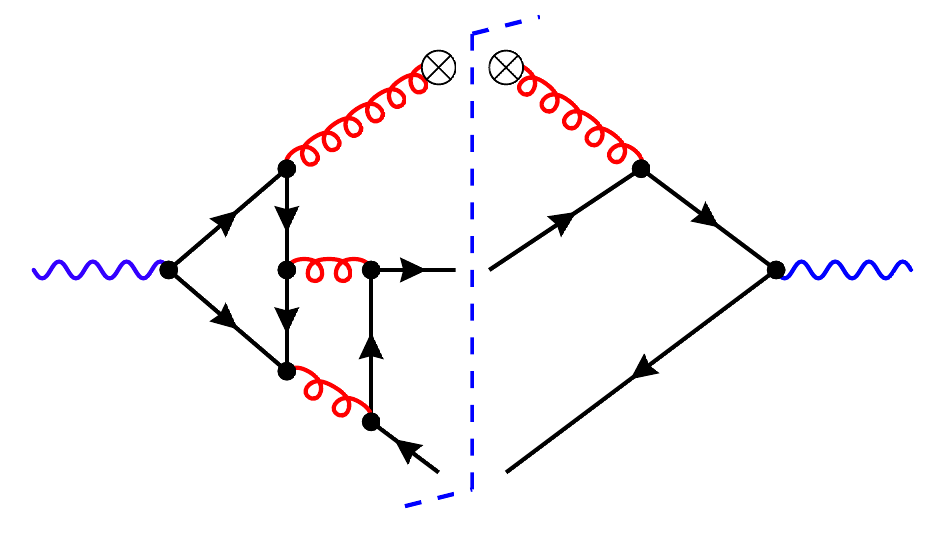}
        \caption{VVR}
        \label{fig:subfig3}
    \end{subfigure}
       \hfill
    \begin{subfigure}{0.48\linewidth}
        \centering
        \includegraphics[width=\linewidth]{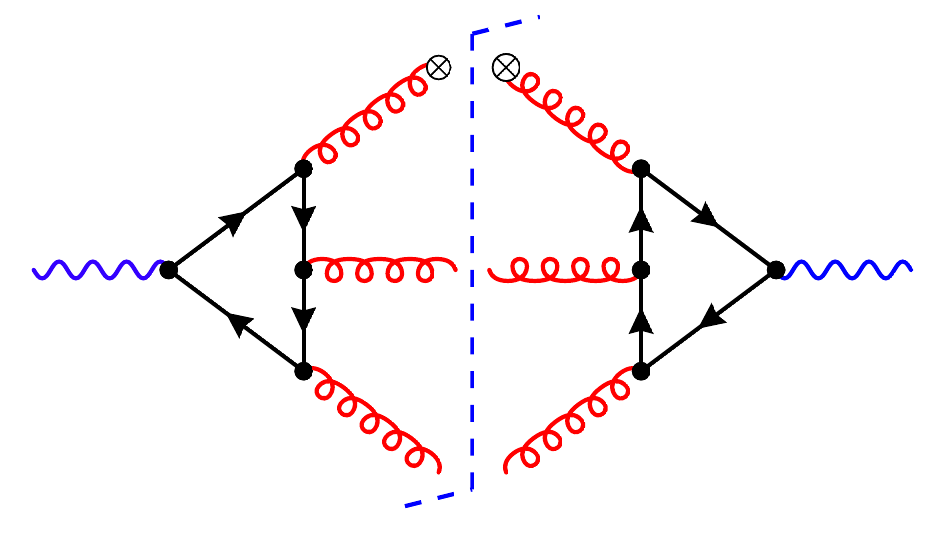}
        \caption{VV$^*$R}
        \label{fig:subfig4}
    \end{subfigure}
    \hfill
    \begin{subfigure}{0.48\linewidth}
        \centering
        \includegraphics[width=\linewidth]{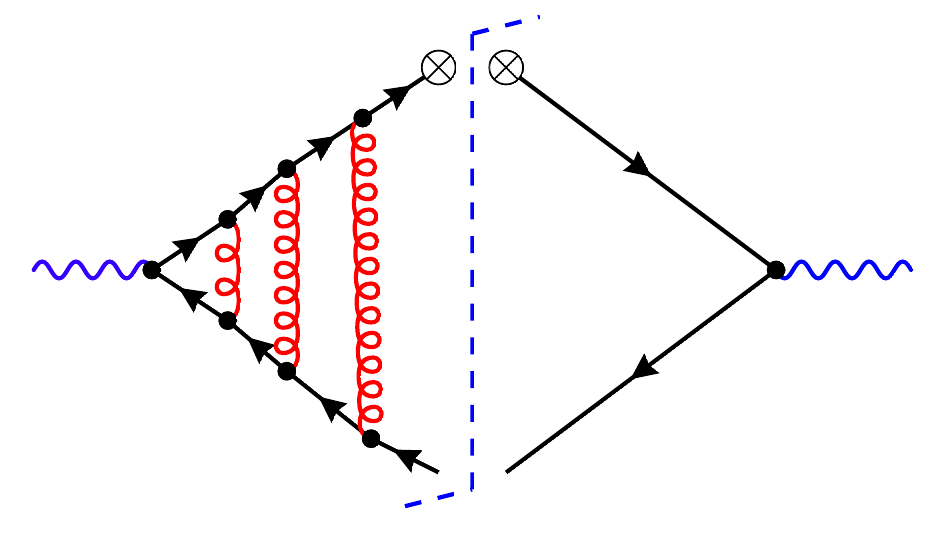}
        \caption{VVV}
        \label{fig:subfig5}
    \end{subfigure}
    \caption{Sample cut Feynman diagrams contributing to SIA coefficient functions, where the crossed bubble represents a single identified parton.}
    \label{fig:sampleFeynman}
\end{figure}

\noindent{\it The method.--} At N$^3$LO, one, in principle, needs to compute the contributions from all relevant cuts, i.e., triple real (RRR), double-real virtual (VRR), double-virtual real (VVR), virtual-squared real (VV$^*$R) and triple virtual (VVV). Some sample Feynman diagrams are shown in Fig.~\ref{fig:sampleFeynman}. Although the contribution from VVV has been known for over a decade~\cite{Baikov:2009bg,Lee:2010cga,Gehrmann:2010ue}, the real corrections are significantly more complex and have never been evaluated, preventing the extraction of SIA at N$^3$LO. 

Fortunately, an exploration of analytic continuation from deep inelastic scattering (DIS) could significantly simplify the computations for SIA. At LO, the partonic channel for DIS reads $\gamma^* (q) + \text{quark} (p) \to \text{quark} (k)$, where the corresponding Bjorken variable in space-like region is $x_B = -q^2/(2 p \cdot q)$ with $q^2<0$. 
While for SIA, the corresponding partonic channel reads $\gamma^* (q) \to  \text{anti-quark} (p)+  \text{quark} (k)$, where $x = 2 p \cdot q/q^2$ with $q^2>0$. These two processes are related by crossing symmetry, i.e., SIA can be obtained from DIS by replacing an incoming quark with an outgoing anti-quark. In terms of momenta and Bjorken variables, the analytic continuation rule reads
\begin{align}
      \label{eq:acRule}
p\to -p\,, \quad x_B = \frac{-q^2}{2 p\cdot q }  \to \frac{-q^2}{-2 p \cdot q }  = \frac{1}{x}\,.
\end{align}
The above rule has been noted long ago in Refs.~\cite{Drell:1969jm,Drell:1969wd}. However, a direct analytic continuation for cross-section-level quantities fails beyond two-loop order~\cite{Mitov:2006ic,Moch:2007tx,Almasy:2011eq,Luo:2019bmw}. This problem has been carefully analyzed and resolved in Ref.~\cite{Chen:2020uvt} within the context of analytic continuation for transverse-momentum-dependent beam and fragmentation functions at N$^3$LO. We extend this method to scattering processes in full QCD and further address the previously unconsidered analytic continuation for the VV$^*$R contribution. The issue stems from the fact that analytic continuation should be performed at the bare amplitude level, i.e., before summing over the corresponding complex conjugate contributions, rather than at the renormalized cross-section level.
 
The analytic continuation rule $x_B\to 1/x$ only crosses the branch cut $x_B=x=1$. Therefore, it is sufficient to analyze the threshold limits for all different cut contributions at the bare amplitude level. Tab.~\ref{tab:DIS_limit} and Tab.~\ref{tab:SIA_limit} show the structures in the threshold limit for DIS and SIA, respectively. In the tables, the results are organized to factor out an overall factor $\big[(p+q)^2\big]^{-3 \epsilon}$ for DIS and $\big[(-p+q)^2\big]^{-3 \epsilon} $ for SIA. The first factor gives rise to the prescription for $x_B$, i.e $(p+q)^2+ i 0^+$ gives $x_B = \text{Re}\left[x_B\right] - i0^+$, where ``Re" denotes the real part. These two factors are real in their respective regions and are related by the crossing rule $p \to -p$, such that the remaining parts are related to each other by $1-x_B \to (1/x-1)e^{i \pi }$~\cite{Chen:2020uvt}. Special attention is required for the analytic continuation of the VV$^*$R contribution, since it simultaneously involves V and its complex conjugate. To this end, we carefully track the contributions from V and V$^*$ by using $x_B$ and it's complex conjugate $x_B^*$, respectively. Here, $x_B$ and $x_B^*$ are $x_B = \text{Re}\left[x_B\right] - i 0^+$ and $x_B^* = \text{Re}\left[x_B\right] + i 0^+$. The analytic continuation rule for $x^*_B$ then reads $1-x_B^* \to (1/x^*-1) e^{-i\pi}$. As a brief summary, the analytic continuation rule in Eq.~\eqref{eq:acRule} is modified to
\begin{align}
\label{eq:aCMod}
    p\to & -p\,, \quad 1-x_B = (1/x-1)e^{i \pi} \,, \nonumber \\
    & 1-x_B^* \to (1/x^*-1) e^{-i\pi}\,.
\end{align}

\begin{widetext}

\begin{table}[ht]
\centering
\begin{tabular}{c|c}
  \hline Cuts &  Threshold-limit structures for DIS \\ 
 \hline
   RRR  &  $\big[(p+q)^2\big]^{-3 \epsilon}  B_0$    \\ 
   VRR & $\big[(p+q)^2\big]^{-3 \epsilon}  \left[ B_{s} e^{i \pi \epsilon} + B_{c} (1-x_B)^{ \epsilon} \right]$    \\  
 VVR & $\big[(p+q)^2\big]^{-3 \epsilon}  \left[ B_{s,s} e^{2i \pi \epsilon} + B_{s,c} e^{i \pi \epsilon}  (1-x_B)^{ \epsilon} + B_{c,c} (1-x_B)^{ 2 \epsilon}\right]$   \\  
 VV$^*$R & $\big[(p+q)^2\big]^{-3 \epsilon}   \left[ B_{s,s}  + B_{c,s} e^{i \pi \epsilon}  (1-x_B^{*})^{ \epsilon} + B_{c,s} e^{-i \pi \epsilon}  (1-x_B)^{ \epsilon} + B_{c,c} (1-x_B)^{ \epsilon} (1-x_B^*)^{\epsilon}\right]$  \\ 
 \hline
\end{tabular}
\caption{The structures of the DIS coefficient function for different cuts in the threshold limit at N$^3$LO, where the coefficients $B_0, \, B_{s}, B_{c,s}, \cdots$ are purely real, and $x_B^*$ is the complex conjugate of $x_B$.}
\label{tab:DIS_limit}
\end{table}
\begin{table}[ht]
\centering
\begin{tabular}{c|c}
\hline  Cuts &  Threshold-limit structures for SIA \\ \hline
   RRR  &  $\big[(-p+q)^2\big]^{-3 \epsilon}  F_0$    \\  
   VRR & $\big[(-p+q)^2\big]^{-3 \epsilon}  e^{i \pi \epsilon} \left[ F_{s} + F_{c} (1-x)^{ \epsilon} \right]$    \\ 
 VVR & $\big[(-p+q)^2\big]^{-3 \epsilon}  e^{2i \pi \epsilon} \left[ F_{s,s} + F_{s,c} (1-x)^{ \epsilon} + F_{c,c} (1-x)^{ 2 \epsilon}\right]$   \\   
 VV$^*$R & $\big[(-p+q)^2\big]^{-3 \epsilon}  \left[ F_{s,s}  + F_{c,s}   (1-x^*)^{ \epsilon} + F_{c,s}   (1-x)^{ \epsilon} + F_{c,c} (1-x)^{ \epsilon} (1-x^*)^{\epsilon}\right]$  \\ 
 \hline 
\end{tabular}
\caption{The structures of SIA coefficient function for different cuts in the threshold limit at N$^3$LO, where the coefficients $F_0, \, F_{s}, F_{c,s}, \cdots$ are purely real, and $x^*$ is the complex conjugate of $x$.}
\label{tab:SIA_limit}
\end{table}

\end{widetext}

The currently known coefficient functions for DIS at N$^3$LO~\cite{Vermaseren:2005qc} are computed from forward scattering, i.e., they are the summation of all cut contributions and their complex conjugates. Tab.~\ref{tab:DIS_limit} shows that the summation of the RRR, VRR, and their complex conjugate contributions still produces correct results under a direct analytic continuation, up to a purely imaginary part. However, this is not the case for VVR and VV$^*$R. Explicitly, 
\begin{align}
\label{eq:vvrDifference}
&\left[ \mathcal{AC}(\text{VVR}) + \text{c.c.} \right] - \text{Re}\left[\mathcal{AC}(\text{VVR}+\text{c.c.})\right] \nonumber \\ 
   & \qquad \propto B_{s,c} \, \sin^2 (\pi \epsilon)  \neq 0\,, \nonumber 
   \\ 
    &\mathcal{AC}\big(\text{VV}^*\text{R}\big)-\text{Re}\left[ \mathcal{AC}\big(\text{VV}^*\text{R}\big|_{x_B^* \to x_B}\big) \right] \nonumber \\
    & \qquad  \propto  \left[ B_{c,c} + B_{c,s} \right] \sin^2(\pi \epsilon) \neq 0 \,, 
\end{align}
where $\mathcal{AC}$ and c.c. denote analytic continuation based on the rule in Eq.~\eqref{eq:aCMod} and complex conjugate, respectively. The $\mathcal{AC}(\text{VV}^*\text{R}\big|_{x_B^* \to x_B})$ represents the direct analytic continuation of the cross-section-level quantity for VV$^*$R. Unlike the cases of RRR and VRR, the above differences are non-zero and contribute to finite part when multiplying with an $\epsilon$-divergent coefficient.

To maximally reduce the complexity of computations, we propose the following analytic continuation formula for the N$^3$LO correction, schematically:
\begin{align}
\label{eq:acFormula}
d &\sigma^{(3),\,{\text{SIA}}}  = \text{Re}\Bigg[\mathcal{AC}\Big({ {d \sigma^{(3),\,\text{DIS}}_{\text{}}}{}} \Big)  \nonumber \\
&  - \mathcal{AC}\Big(d \sigma^{(3),\,\text{DIS}}_{ \text{VVR} } +\text{c.c.}\Big) - \mathcal{AC}\Big(d \sigma^{(3),\,\text{DIS}}_{\text{VV}^*\text{R} }\big|_{x_B^* \to x_B} \Big)   \nonumber \\
& + \left\{\mathcal{AC}\Big({d \sigma^{(3),\,\text{DIS}}_{{\text{VVR}  } }}{}\Big)  +\text{c.c.}\right\}  + \mathcal{AC}\Big({d \sigma^{(3),\,\text{DIS}}_{ {\text{VV}^*\text{R} } } } \Big)   \Bigg]_{\text{}}^{\text{}}.
\end{align}
We emphasize that each term above is understood as an unrenormalized quantity. The first term represents a direct analytic continuation of the differential cross section at N$^3$LO for DIS~\cite{Vermaseren:2005qc}, which includes problematic contributions from VVR and VV$^*$R. In the second line, the problematic contributions are subtracted, and in the third line, the contributions from the correct analytic continuation are restored. Therefore, to obtain the coefficient functions for SIA at N$^3$LO, we need to explicitly compute the VVR and VV$^*$R contributions for DIS. Alternatively, we can compute the VVR and VV$^*$R contributions directly for SIA and analytically continue them back to the DIS region. The VVR and VV$^*$R contributions for both the DIS and SIA regions have not been reported in the literature. In the next section, we explicitly evaluate these contributions in the SIA region.

\noindent{\it Computations of VVR and VV$^*$R.--}
We use the following projectors to extract the transverse and longitudinal partonic coefficient functions~\cite{Rijken:1996ns,Mitov:2006wy}
\begin{align}
\label{eq:proctor}
    C_{T,p}(x,Q^2) &= \frac{1}{d-2} \left(  -\frac{2 p \cdot q}{q^2} g^{\mu \nu} - \frac{2}{p \cdot q} p^\mu p^\nu  \right) \hat{W}_{\mu \nu} \,, \nonumber \\ 
    C_{L,p}^{}(x, Q^2) &= \frac{1}{p\cdot q} p^\mu p^\nu \hat{W}_{\mu \nu}\,, 
\end{align}
where $\hat{W}_{\mu \nu}$ is the parton structure tensor defined by
$\hat{W}_{\mu \nu}(p,q) = \frac{x^{1-2\epsilon}}{4 \pi}  \int d\text{PS}^{(l)} M_\mu^{\gamma^*\to p+X}  (M_\nu^{\gamma^* \to p+X})^*\,.$
Here, the factor $x^{1-2\epsilon}$ is from the phase-space of the single identified parton, $\int d \text{PS}^{(l)}$ is the $l$-body invariant phase space measure, and $M_\mu^{\gamma^* \to p +X}$ is the amplitude for process
$\gamma^*(q) \to p + p_1 + p_2 + \cdots p_l $
with $p$ and $p_1, \cdots, p_l$ being the momenta of the identified parton and unidentified partons, respectively. 

Using the above projectors, the computations for VVR and VV$^*$R follow the standard multi-loop techniques. We utilize \texttt{QGRAF}~\cite{Nogueira:1991ex} to generate all relevant Feynman diagrams, and \texttt{Form}~\cite{Vermaseren:2000nd, Ruijl:2017dtg}, \texttt{Color}~\cite{vanRitbergen:1998pn} to evaluate Dirac and color algebra. The method of reverse unitarity~\cite{ANASTASIOU2002220} is employed to enable the application of standard integration-by-parts (IBP) reductions~\cite{Chetyrkin:1980pr, Chetyrkin:1981qh, Laporta:2000dsw} for mixed loop and phase-space integrals. In this work, we utilize \texttt{FIRE6}~\cite{Smirnov:2019qkx}, \texttt{Kira}~\cite{Klappert:2020nbg} and \texttt{Blade}~\cite{Guan:2024byi} to reduce a large number of integrals to a basis of integrals, called {\it master integrals}. We identified all master integrals from 24 integral families for VVR and from 10 integral families for VV$^*$R. We then directly solve these master integrals using the differential equation (DE) method~\cite{Gehrmann:1999as}, with the boundary constants determined by matching to inclusive integrals for VVR and VV$^*$R~\cite{Magerya:2019cvz, Maheria:2022dsq}. The DE system is derived with the aid of \texttt{LiteRed}~\cite{Lee:2012cn}, and converted into canonical form~\cite{Henn:2013pwa} by the package \texttt{CANONICA}~\cite{Meyer:2017joq}. It should be noted that, under analytic continuation, contributions from different branches at $x=1$, i.e. $(1-x)^{a \epsilon}$ with $a$ being integers, should be separated. Since different branches do not interact before expanding $\epsilon$, it is straightforward to construct and solve the DEs for different branches independently.

The explicit bare result for VVR up to finite term in $\epsilon$ only involves letters $\{ x, 1-x\}$, which can be expressed in term of harmonic polylogarithms (HPLs)~\cite{Remiddi:1999ew}. On the other hand, the bare result for VV$^*$R contains letters $\{x,1-x, 2-x\}$ in finite term, which involves Goncharov multiple polylogarithms (GPLs). We use \texttt{HPL}~\cite{Maitre:2005uu} and \texttt{Polylogtools}~\cite{Duhr:2019tlz} packages to deal with HPLs and GPLs, respectively. 

\noindent{\it Result.--} In addition to the new bare results for VVR and VV$^*$R, the lower-order results in $a_s$ up to higher-order in $\epsilon$ and collinear counterterms are also necessary. We explicitly computed NNLO corrections up to $\mathcal{O}(\epsilon^2)$ for both DIS and SIA, which are not yet publicly available in the literature, and verified that the analytic continuation relation holds between them. The collinear counterterms for FFs are composed of NNLO time-like splitting functions~\cite{Mitov:2006ic,Moch:2007tx,Almasy:2011eq,Chen:2020uvt}. Notice that the collinear counterterms for parton distribution functions (PDFs), which are composed of NNLO space-like splitting functions~\cite{Moch:2004pa,Vogt:2004mw}, are also required to extract the bare results at N$^3$LO for DIS from the renormalized ones~\cite{Vermaseren:2005qc}. By utilizing the analytic continuation formula in Eq.~\eqref{eq:acFormula}, and performing mass factorization using lower-order results and collinear counterterms, we managed to obtain the analytic results for SIA partonic coefficient functions at N$^3$LO, except for $\delta(1-x)$ term. The analytic continuation of $\delta(1-x)$ term requires to track terms $(1-x_B)^{-1+ a \, \epsilon}$ in all cut contributions as well as $\delta(1-x_B)$ term from the analytic 3-loop form factor~\cite{Lee:2010cga,Gehrmann:2010ue}, which is quite involved. Instead, we determine the $\delta(1-x)$ contribution through sum rules, i.e., integrating over $x$ and match with the inclusive cross section at N$^3$LO~\cite{GORISHNY1991144,PhysRevLett.66.560}. 

The final results for coefficient functions are expressed in terms of HPLs, and we have performed several consistency checks:
First, all poles in $\epsilon$ cancel when using the three-loop time-like splitting functions~\cite{Mitov:2006ic,Moch:2007tx,Almasy:2011eq, Chen:2020uvt}. Second, the $\delta(1-x)$ term at N$^3$LO matches precisely with results derived from threshold resummation at both N$^3$LL and N$^4$LL accuracy~\cite{Moch:2009my,Xu:2024rbt}. Moreover, for the logarithmically enhanced terms at leading, next-to-leading power and beyond in the threshold limit, we find exact agreement with earlier predictions~\cite{Blumlein:2006pj,Moch:2009hr,AH:2020xll}. These agreements not only strongly support the correctness of our results, but also confirm the validity of the factorization formula at both leading and next-to-leading power as established in previous literature. Third, by integrating over $x$ in our explicit results for VVR and VV$^*$R, we find complete agreement with the inclusive results in Ref.~\cite{Jakubcik:2022zdi}.
Fourth, the NNLO corrections are in full agreement with Refs.~\cite{Rijken:1996ns,Mitov:2006wy}. The final results are too lengthy to be presented here, we have included them in an ancillary file submitted with this paper.

\begin{figure}[t]
	\centering	\includegraphics[width=0.5\textwidth]{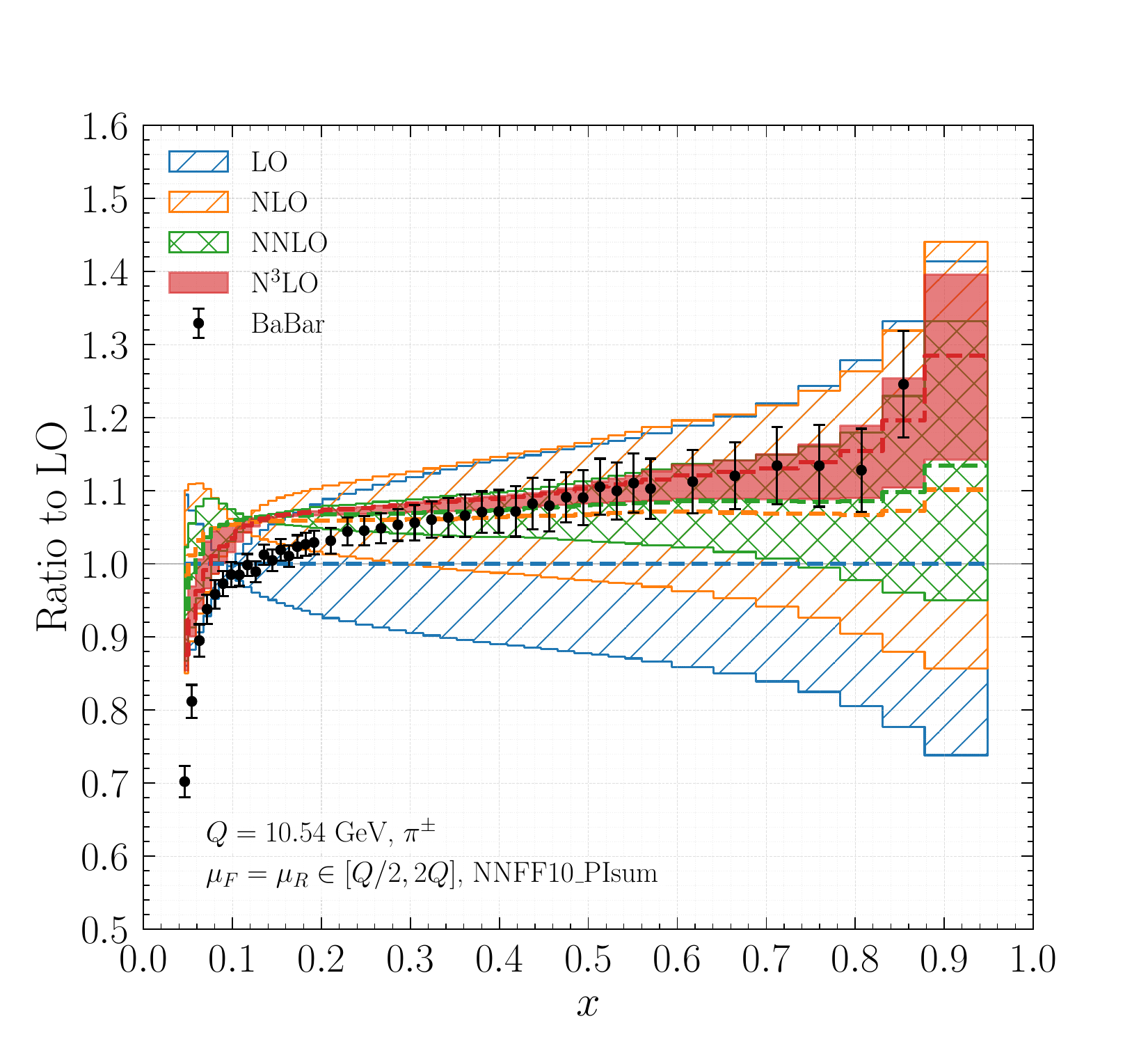}
\caption{
Comparison of BaBar data with theoretical predictions at each perturbative order, normalized to the LO results with central scale $\mu_F = \mu_R = Q$. }
	\label{fig-combine}
\end{figure}

To assess the phenomenological impact of our results, we perform numerical calculations at N$^3$LO. Specifically, we provide predictions for the $\theta$-inclusive case in Eq.~\eqref{eq:siaDiff}, corresponding to the sum of the longitudinal and transverse differential cross sections. For the numerical evaluation of HPLs, we utilize the package~\texttt{hplog}~\cite{Gehrmann:2001pz}. As a benchmark comparison, we confront our predictions with the high precision BaBar measurements at collision energy $\sqrt{s}=10.54$ GeV~\cite{BaBar:2013yrg}. 
In this work, we employ the NNFF1.0~\cite{Bertone:2017tyb} parametrization, which provides FFs from LO to NNLO accuracy. For partonic coefficient functions up to NNLO, we employ the corresponding FFs at the same order. Since FFs at N$^3$LO accuracy are not yet available, we use the NNLO FFs as a proxy for N$^3$LO coefficient functions to illustrate the impact of N$^3$LO corrections. The results are shown in Fig.~\ref{fig-combine}, which displays the detailed comparison between theoretical predictions at each perturbative order and BaBar data, normalized to the LO results with central scale $\mu_F = \mu_R = Q$.

We begin by analyzing the scale uncertainties of our results, represented by the colored bands, which arise from varying the QCD scales simultaneously by a factor of two. The figure clearly shows a substantial reduction in scale uncertainties as the perturbative order increases. Notably, our N$^3$LO results exhibit up to a factor of two smaller uncertainties compared to NNLO, and we anticipate further improvement once the N$^3$LO accurate FFs become available. When compared to the experimental uncertainties from BaBar, indicated by error bars in the figure, our results demonstrate the importance of achieving N$^3$LO accuracy to meet the precision of existing experimental data. 

To quantify the impact of higher-order pQCD corrections, we further analyze the standard K-factor, which is defined as the ratio of the central-scale cross sections at different perturbative orders to the LO result. These K-factor, represented by the colored dashed lines in Fig.~\ref{fig-combine}, reveal increasingly significant contributions from higher order terms near the phase-space boundaries for both small-$x$ and large-$x$ regions. Notably, the N$^3$LO correction can reach about 30$\%$ for $x\sim 0.9$. Such a substantial correction is essential to achieve agreement with the BaBar data.

\noindent{\it Conclusion.--}
In this work, we present the first analytical computation of QCD corrections to SIA at N$^3$LO. By employing analytic continuation from DIS and explicitly evaluating the VVR and VV$^*$R contributions, we derive the N$^3$LO coefficient functions for the transverse and longitudinal cross sections, encapsulated in Eq.~\eqref{eq:factorization}. Rigorous validation through pole cancellation, consistency with inclusive cross sections at N$^3$LO, and agreement with threshold resummation at N$^4$LL accuracy confirm the robustness of our results.  

Phenomenologically, the N$^3$LO predictions exhibit a marked reduction in scale uncertainties, by up to a factor of two compared to NNLO, and significantly improve the description of BaBar experimental data. The corrections are particularly pronounced in the large-$x$ region, reaching about $30\%$ for $x \sim 0.9$. These advancements underscore the critical role of higher-order perturbative corrections in precision QCD and provide a refined framework for extracting FFs.  

Looking ahead, a global determination of N$^3$LO FFs based on SIA data will become feasible once the full set of four-loop time-like splitting functions are available. This will further enhance the precision of hadronization studies and enable tighter constraints on non-perturbative QCD dynamics. The achieved accuracy highlights the necessity of N$^3$LO computations in bridging the gap between theoretical predictions and experimental precision in strong interaction physics.

\noindent{\it Note added.--} 
As we completed this work, we became aware of a related study~\cite{Magerya:2025qgp}, which calculates the relevant master integrals for SIA at N$^3$LO.

%======================================================================
\noindent{\it Acknowledgements.--}
We thank T. Gehrmann to provide us the Fortran package \texttt{hplog}~\cite{Gehrmann:2001pz}. We also thank P. Garbarino, L. Bonino, M. L\"ochner, K. Sch\"onwald, V. Sotnikov, Z. Xu and S. Zoia for useful discussions. The Feynman diagrams are drawn with the aid of \texttt{FeynGame}~\cite{Bundgen:2025utt}. This work is supported by the National Natural Science Foundation of China under Grants Nos.~12475139, 12425505 and 12525508, and by the Swiss National Science Foundation (SNF) under contract 200020-204200 and the European Research Council (ERC) under the European Union's Horizon 2020 research and innovation programme grant agreement 101019620 (ERC Advanced Grant TOPUP).

\noindent{\it Data availability.--} The data that support the findings of this Letter are openly available~\cite{he2025}, embargo periods may apply.

% He, C.-. qi ., Xing, H., Yang, T.-Z., & Zhu, H. (2025). Ancillary files for single-inclusive hadron production in electron-positron annihilation at next-to-next-to-next-to-leading order in QCD [Data set]. Zenodo. https://doi.org/10.5281/zenodo.15206524

%======================================================================
% \appendix
% \section{Appendix A}
%======================================================================
\bibliographystyle{apsrev4-1}
\bibliography{SIAN3LO}
\end{document}